\documentclass[a4paper,twocolumn]{esapub2005} 
\usepackage{epsf}
\usepackage{times}
\usepackage{natbib}

\newcommand\aj{AJ}
\newcommand\apj{ApJ}
\newcommand\apjl{ApJ}
\newcommand\aap{A\&A}
\newcommand\mnras{MNRAS}
\newcommand\pasp{PASP}
\newcommand\grl{Geophys.~Res.~Lett.}

\title{Calibrating the Solar Dynamo: \hspace{3.0in} Magnetic Activity Cycles of Southern Sun-like Stars}
\author[1]{T.~S.~Metcalfe}
\affil[1]{High Altitude Observatory, NCAR, P.O. Box 3000, Boulder, CO 80307 USA (travis@ucar.edu, knoe@ucar.edu)}
\author[2]{T.~J.~Henry}
\affil[2]{Department of Physics and Astronomy, Georgia State University, Atlanta, GA 30302 USA (thenry@chara.gsu.edu)}
\author[1]{M.~Kn{\"o}lker}
\author[3]{D.~R.~Soderblom}
\affil[3]{Space Telescope Science Institute, 3700 San Martin Drive, Baltimore, MD 21218 USA (drs@stsci.edu)}

\begin{document}


\maketitle

\begin{abstract}

The solar magnetic activity cycle is responsible for periodic episodes of 
severe space weather, which can perturb satellite orbits, interfere with 
communications systems, and bring down power grids. Much progress has 
recently been made in forecasting the strength and timing of this 11-year 
cycle, using a predictive flux-transport dynamo model \cite{dik05,dik06}. 
We can strengthen the foundation of this model by extending it to match 
observations of similar magnetic activity cycles in other Sun-like stars, 
which exhibit variations in their Ca~{\sc ii} H and K emission on time 
scales from 2.5 to 25 years \cite{bal95}. This broad range of cycle 
periods is thought to reflect differences in the rotational properties and 
the depth of the surface convection zone for stars with various masses and 
ages. Asteroseismology is now yielding direct measurements of these 
quantities for individual stars, but the most promising asteroseismic 
targets are in the southern sky ($\alpha$~Cen~A, $\alpha$~Cen~B, 
$\beta$~Hyi), while the existing activity cycle survey is confined to the 
north. We are initiating a long-term survey of Ca~{\sc ii} H and K 
emission for a sample of 92 southern Sun-like stars to measure their 
magnetic activity cycles and rotational properties, which will ultimately 
provide independent tests of solar dynamo models.

\end{abstract}


\section{Context} 

Astronomers have been making telescopic observations of sunspots since the 
time of Galileo, gradually building a historical record showing a periodic 
rise and fall in the number of sunspots every 11 years. We now know that 
sunspots are regions with an enhanced local magnetic field, so this 
11-year cycle actually traces a variation in surface magnetism. Attempts 
to understand this behavior theoretically often invoke a combination of 
differential rotation, convection, and meridional flow to modulate the 
field through a magnetic dynamo. Significant progress in this area quickly 
followed after helioseismology allowed the detailed characterization of 
the Sun's interior structure and dynamics over the past decade (e.g., the 
convection zone depth and meridional flow).

\begin{figure}
\epsfxsize 3.0in
\epsffile{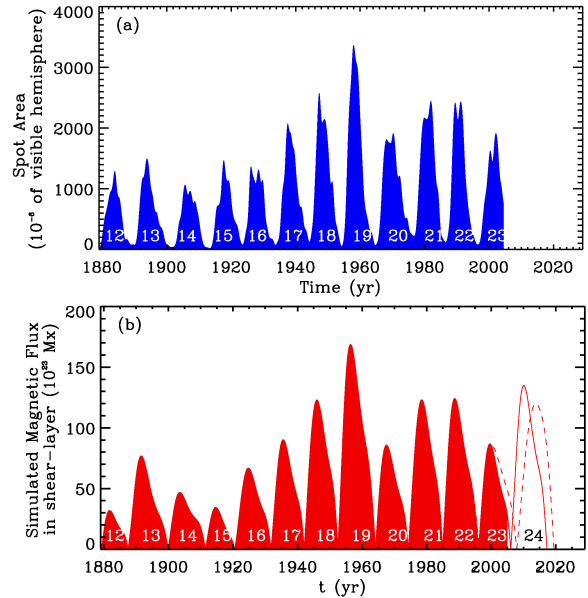}
\caption{A flux-transport dynamo model correctly reproduces the relative 
amplitudes of the past 8 solar cycles, and then predicts the strength and 
timing of the next cycle (from \cite{dik06}, reproduced with permission).}
\end{figure}

This has gradually led to the development of a predictive flux-transport 
dynamo model, which has recently been used to forecast the timing 
\cite{dik05} and strength \cite{dik06} of the upcoming solar magnetic 
cycle. By calibrating the relationship between sunspot area and surface 
magnetic flux using modern measurements, historical sunspot data going 
back more than 125 years could be used to test the model. After 
initializing the dynamo with the first several cycles, the model correctly 
reproduced the relative amplitudes of the subsequent 8 cycles, and was 
then used to predict the characteristics of future cycles (see Fig.~1). 
This is not merely of academic interest, since the magnetic maximum of the 
Sun is accompanied by severe space weather events. Advance knowledge of 
the timing and overall strength of the coming cycle allows us to prepare 
for the adverse consequences, which include the destabilization of 
orbiting satellites, interference with communications systems, and the
disruption of power grids.

While the success of this dynamo model in reproducing the solar cycle is 
very encouraging, there will soon be new opportunities for independent 
tests using the magnetic activity cycles of other Sun-like stars. The 
properties of the dynamo---most notably the cycle period---are sensitive 
to specific physical circumstances, such as the meridional flow, the 
degree of latitudinal differential rotation and the depth of the 
convection zone. By matching the dynamo model to the single set of 
circumstances represented by the Sun, we risk the possibility of fine 
tuning, which could ultimately limit our comprehension of the underlying 
physical mechanisms. If we exploit the magnetic activity cycles observed 
in a wide variety of Sun-like stars, we can extend the calibration to 
dozens of unique circumstances and ensure that the physical basis of the 
model is more broadly applicable. Specifically, observations of stellar 
activity cycle periods and differential rotation profiles (from this 
proposed survey, and its existing northern counterpart) as well as stellar 
convection zone depths (from ground- and space-based asteroseismology) 
will allow us to test the dynamo model by inverting the implied stellar 
meridional flow, and comparing it to the solar case.

Although we can rarely observe spots on other Sun-like stars directly, 
these areas of concentrated magnetic field produce strong emission in the 
Ca~{\sc ii}~H (396.8~nm) and K (393.4~nm) spectral lines. The intensity of 
the emission scales with the amount of non-thermal heating in the 
chromosphere, making these lines a useful spectroscopic proxy for the 
strength of, and fractional area covered by, magnetic fields \cite{lei59}. 
Disk-integrated time-series measurements of Ca~{\sc ii}~H and K emission 
in the Sun easily reveal the 11-year cycle (see Fig.~2). Similar 
measurements for a large sample of northern Sun-like stars began at the 
Mount Wilson Observatory in 1966 \cite{wil78}, and continue to this day 
\cite{bal06}.

\begin{figure}[t]
\epsfxsize 3.0in
\epsffile{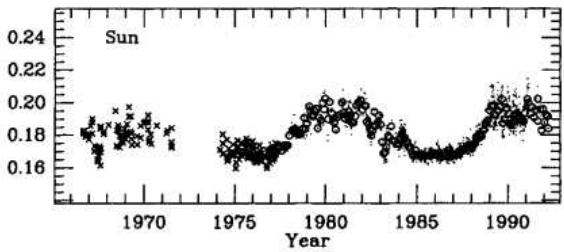}
\caption{The Ca~{\sc ii}~HK index of the Sun observed by the Mount Wilson 
survey (adapted from \cite{bal95}).}
\end{figure}

The Mount Wilson Ca~HK survey, as it is known, has revealed magnetic 
activity variations in Sun-like stars with cycle periods ranging from 2.5 
to more than 25 years. The frequent time sampling for 111 stars since 1980 
also allows the measurement of rotation periods, from the repeated 
signature of individual active regions rotating into view and then out 
again. In some cases the rotation periods appear to drift over longer 
timescales, due to the migration of active regions to lower latitudes 
which are rotating faster (the same mechanism that is responsible for the 
``Butterfly Diagram'' on the Sun). Thus, it is sometimes possible to 
measure or place limits on the degree of latitudinal differential rotation 
in other Sun-like stars. Early analysis of the Mount Wilson data revealed 
an empirical correlation between the mean level of magnetic activity and 
the rotation period normalized by the convective timescale \cite{noy84a}. 
The data also established a relation between the rotation rate and the 
period of the observed activity cycle, which generally supports a dynamo 
interpretation \cite{noy84b}.

It was several decades before similar observations were conducted for a 
large sample of Sun-like stars in the southern sky \cite{hen96}. These 
data confirmed the bimodal distribution in mean stellar activity {\it 
levels} first seen in the northern sample \cite{vp80}, but they could say 
nothing about activity {\it cycles} because the data were confined to a 
single epoch of observation. Another decade has passed, and there is still 
no time-domain survey of Ca~HK emission in southern Sun-like stars. This 
is particularly disappointing in light of the fact that the most promising 
candidates for asteroseismology, which will soon allow the direct 
measurement of stellar convection zone depths (see Fig.~3), are stars in 
the southern sky---most notably the brightest solar-type stars 
$\alpha$~Cen~A, $\alpha$~Cen~B, and $\beta$~Hyi which has long been 
studied as a ``future Sun''.

\begin{figure*}[t]
\epsfxsize 5.5in
\hskip 0.5in
\epsffile{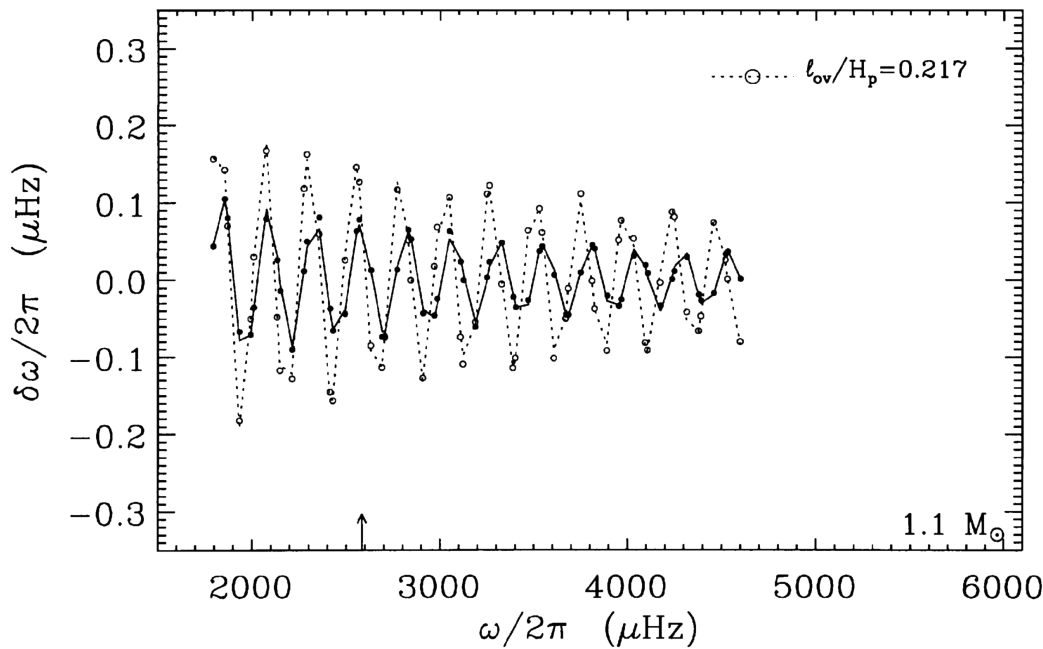}
\caption{The expected oscillation signal from the edge of the convective 
envelope of a 1.1 M$_{\odot}$ star with (dotted) and without (solid) 
overshoot (adapted from \cite{mon00}). Data with this level of 
precision will soon emerge from space-based asteroseismology missions
such as CoRoT and Kepler.}
\end{figure*}

Solar-like oscillations have already been detected in these stars 
\cite{bed04,kje05,bed01}, but tests of solar dynamo models will not be 
possible until their magnetic activity cycle periods are established. 
There is some evidence from sparse x-ray observations of $\alpha$~Cen~A 
that it may exhibit an activity cycle as short as 3.5~years \cite{rob05}, 
despite the fact that it is only 10\% more massive than the Sun and just 
slightly older. A firm measurement of the cycle period in $\alpha$~Cen~A, 
particularly when combined with forthcoming asteroseismic data, could 
provide a very stringent second test of the flux-transport dynamo model.


\section{Project Goals} 

We are proposing to initiate a long-term survey of Ca~{\sc ii}~H and K 
emission for a sample of southern Sun-like stars, establishing a southern 
counterpart of the Mount Wilson Ca~HK survey.

During at least the first two years, we will monitor the sample with 
sufficient frequency (several times per week) to measure the rotation 
periods, and to place limits on the degree of latitudinal differential 
rotation. For those stars with the shortest activity cycles, we will 
either directly measure or provide firm upper limits on the cycle period. 
By comparing our observations to those from the single-epoch survey in 
1996, we will also establish interesting limits on the cycle period for 
those stars with the slowest variations in activity. These initial 
results, particularly the analysis of $\alpha$~Cen~A, will provide a 
compelling justification for the continuation of the survey through 
external funding.

Our longer-term goal, which extends beyond the two-year period covered by 
our proposal, is to position NCAR as a leader in the observation and 
interpretation of magnetic activity cycles, complementing the role of the 
Mount Wilson Observatory in the north. We will constitute a direct link to 
facilitate the use of magnetic activity cycles and asteroseismic data to 
calibrate solar dynamo models, bridging a gap between the solar and 
stellar communities. At the very least, we plan to measure the activity 
cycle periods and rotational properties for a complete sample of 92 
southern Sun-like stars, but we also hope eventually to upgrade the 
instrumentation and enlarge the sample as resources allow.


\section{Methodology} 

The single-epoch survey of Henry et al.~(1996) contained a total of 1016 
observations of 815 individual stars with visual magnitudes between 0.0 
and about 9.0, which were observed using the {\it RC Spec} instrument on 
the CTIO 1.5m telescope. The full survey was completed in 11 nights of 
observing time with typical integration times between 1-10 minutes, 
implying an average data acquisition rate of about 90 observations per 
night. Several sub-samples were defined, including the ``Best \& 
Brightest'' (B) and ``Nearby'' (N) samples, which together contain 92 
individual stars with visual magnitudes between 0.0 and 7.9, and B$-$V 
colors that are approximately solar (see Table~1). All three of the most 
promising southern asteroseismic targets ($\alpha$~Cen~A, $\alpha$~Cen~B, 
and $\beta$~Hyi) are included in this combined (B+N) sample. Since it is a 
full magnitude brighter than the complete sample, integration times will 
often be shorter---but the overhead of slewing the telescope between more 
widely dispersed targets will require more time between observations. 
Thus, we can expect a data acquisition rate roughly comparable to that 
achieved during the 1996 survey.

\begin{table*}
\begin{center}
\footnotesize
\caption{Characteristics of the 92 stars in the survey sample.}\vspace{1em}
\renewcommand{\arraystretch}{1.2}
\begin{tabular}[t]{lccccc|lccccc}
\hline
HD     &  RA(2000)  &  Dec(2000)  &   V   & B$-$V & $\log R^{'}_{HK}$ &
HD     &  RA(2000)  &  Dec(2000)  &   V   & B$-$V & $\log R^{'}_{HK}$ \\
\hline
1273   & 00~16~53.9 & $-$52~39~04 & 6.840 & 0.655 &          & 128620 & 14~39~36.5 & $-$60~50~02 & 0.010 & 0.710 & $-$5.002 \\
1581   & 00~20~04.3 & $-$64~52~29 & 4.226 & 0.576 & $-$4.839 & 128621 & 14~39~35.1 & $-$60~50~14 & 1.350 & 0.900 & $-$4.923 \\
2151   & 00~25~45.1 & $-$77~15~15 & 2.820 & 0.618 & $-$4.996 & 130948 & 14~50~15.8 & $+$23~54~43 & 5.863 & 0.576 &          \\
3443   & 00~37~20.7 & $-$24~46~02 & 5.572 & 0.715 & $-$4.893 & 131156 & 14~51~23.4 & $+$19~06~02 & 4.700 & 0.730 & $-$4.323 \\
4308   & 00~44~39.3 & $-$65~38~58 & 6.546 & 0.655 &          & 131977 & 14~57~28.0 & $-$21~24~56 & 5.723 & 1.024 & $-$4.484 \\
4628   & 00~48~23.0 & $+$05~16~50 & 5.742 & 0.890 & $-$4.854 & 136352 & 15~21~48.1 & $-$48~19~03 & 5.652 & 0.639 &          \\
7570   & 01~15~11.1 & $-$45~31~54 & 4.959 & 0.571 &          & 136466 & 15~22~36.5 & $-$47~55~16 & 7.685 & 0.722 &          \\
10360  & 01~39~47.7 & $-$56~11~34 & 5.900 & 0.800 & $-$4.753 & 140538 & 15~44~01.8 & $+$02~30~55 & 5.865 & 0.684 &          \\
10361  & 01~39~47.2 & $-$56~11~44 & 5.800 & 0.860 & $-$4.881 & 140901 & 15~47~29.1 & $-$37~54~59 & 6.010 & 0.715 &          \\
10476  & 01~42~29.8 & $+$20~16~07 & 5.242 & 0.836 & $-$4.930 & 141004 & 15~46~26.6 & $+$07~21~11 & 4.422 & 0.604 &          \\
10700  & 01~44~04.1 & $-$15~56~15 & 3.495 & 0.727 & $-$4.962 & 147513 & 16~24~01.3 & $-$39~11~35 & 5.385 & 0.625 &          \\
14412  & 02~18~58.5 & $-$25~56~44 & 6.343 & 0.724 &          & 147584 & 16~28~28.1 & $-$70~05~04 & 4.900 & 0.555 &          \\
16160  & 02~36~04.9 & $+$06~53~13 & 5.791 & 0.918 & $-$4.847 & 153631 & 17~01~10.8 & $-$13~34~02 & 7.124 & 0.608 &          \\
17051  & 02~42~33.5 & $-$50~48~01 & 5.400 & 0.561 &          & 155885 & 17~15~20.9 & $-$26~36~09 & 5.110 & 0.860 & $-$4.569 \\
20407  & 03~15~06.4 & $-$45~39~53 & 6.762 & 0.586 &          & 155886 & 17~15~20.9 & $-$26~36~09 & 5.070 & 0.850 & $-$4.577 \\
20766  & 03~17~46.2 & $-$62~34~31 & 5.529 & 0.641 & $-$4.646 & 156026 & 17~16~13.4 & $-$26~32~46 & 6.327 & 1.144 & $-$4.447 \\
20794  & 03~19~55.7 & $-$43~04~11 & 4.260 & 0.711 & $-$4.977 & 156274 & 17~19~03.8 & $-$46~38~10 & 5.330 & 0.770 & $-$4.941 \\
20807  & 03~18~12.8 & $-$62~30~23 & 5.239 & 0.600 & $-$4.787 & 156384 & 17~18~57.2 & $-$34~59~23 & 5.910 & 1.082 & $-$4.721 \\
21411  & 03~26~11.1 & $-$30~37~04 & 7.877 & 0.716 &          & 158614 & 17~30~23.8 & $-$01~03~47 & 5.314 & 0.715 & $-$4.972 \\
22049  & 03~32~55.8 & $-$09~27~30 & 3.726 & 0.881 & $-$4.469 & 160691 & 17~44~08.7 & $-$51~50~03 & 5.127 & 0.694 &          \\
26965  & 04~15~16.3 & $-$07~39~10 & 4.426 & 0.820 &          & 163840 & 17~57~14.3 & $+$23~59~45 & 6.301 & 0.642 &          \\
30003  & 04~40~17.8 & $-$58~56~38 & 6.529 & 0.677 &          & 165185 & 18~06~23.7 & $-$36~01~11 & 5.900 & 0.615 &          \\
30495  & 04~47~36.3 & $-$16~56~04 & 5.491 & 0.632 &          & 165341 & 18~05~27.3 & $+$02~30~00 & 4.026 & 0.860 & $-$4.637 \\
32778  & 05~02~17.1 & $-$56~04~50 & 7.023 & 0.636 &          & 165401 & 18~05~37.5 & $+$04~39~26 & 6.800 & 0.610 &          \\
37655  & 05~38~01.9 & $-$42~57~49 & 7.434 & 0.600 &          & 165499 & 18~10~26.2 & $-$62~00~08 & 5.473 & 0.592 &          \\
43834  & 06~10~14.5 & $-$74~45~11 & 5.080 & 0.714 &          & 172051 & 18~38~53.4 & $-$21~03~07 & 5.858 & 0.673 &          \\
53705  & 07~03~57.3 & $-$43~36~29 & 5.559 & 0.624 &          & 177565 & 19~06~52.5 & $-$37~48~38 & 6.154 & 0.705 &          \\
59468  & 07~27~25.5 & $-$51~24~09 & 6.726 & 0.694 &          & 178428 & 19~07~57.3 & $+$16~51~12 & 6.086 & 0.702 &          \\
63077  & 07~45~35.0 & $-$34~10~21 & 5.363 & 0.589 &          & 189567 & 20~05~32.8 & $-$67~19~15 & 6.078 & 0.648 &          \\
65907  & 07~57~46.9 & $-$60~18~11 & 5.595 & 0.573 &          & 190067 & 20~02~34.2 & $+$15~35~31 & 7.166 & 0.714 &          \\
67458  & 08~07~00.5 & $-$29~24~11 & 6.798 & 0.600 &          & 190248 & 20~08~43.6 & $-$66~10~55 & 3.554 & 0.751 & $-$4.999 \\
71334  & 08~25~49.5 & $-$29~55~50 & 7.797 & 0.643 &          & 190406 & 20~04~06.2 & $+$17~04~13 & 5.788 & 0.600 &          \\
73524  & 08~37~20.0 & $-$40~08~52 & 6.534 & 0.598 &          & 191408 & 20~11~11.9 & $-$36~06~04 & 5.315 & 0.868 & $-$4.988 \\
76151  & 08~54~17.9 & $-$05~26~04 & 6.000 & 0.661 & $-$4.691 & 194640 & 20~27~44.2 & $-$30~52~04 & 6.612 & 0.724 &          \\
88742  & 10~13~24.7 & $-$33~01~54 & 6.377 & 0.592 &          & 196761 & 20~40~11.8 & $-$23~46~26 & 6.363 & 0.719 &          \\
98231  & 11~18~10.9 & $+$31~31~45 & 3.786 & 0.606 &          & 199288 & 20~57~40.1 & $-$44~07~46 & 6.516 & 0.587 &          \\
102365 & 11~46~31.1 & $-$40~30~01 & 4.892 & 0.664 &          & 201091 & 21~06~53.9 & $+$38~44~58 & 5.200 & 1.069 & $-$4.300 \\
102438 & 11~47~15.8 & $-$30~17~11 & 6.481 & 0.681 &          & 202940 & 21~19~45.6 & $-$26~21~10 & 6.558 & 0.737 & $-$4.870 \\
114260 & 13~09~42.5 & $-$22~11~33 & 7.356 & 0.718 &          & 206860 & 21~44~31.3 & $+$14~46~19 & 5.945 & 0.587 &          \\
114613 & 13~12~03.2 & $-$37~48~11 & 4.849 & 0.693 &          & 207129 & 21~48~15.8 & $-$47~18~13 & 5.579 & 0.601 &          \\
115383 & 13~16~46.5 & $+$09~25~27 & 5.209 & 0.585 &          & 209100 & 22~03~21.7 & $-$56~47~10 & 4.688 & 1.056 & $-$4.559 \\
115617 & 13~18~24.3 & $-$18~18~40 & 4.739 & 0.709 & $-$4.963 & 211415 & 22~18~15.6 & $-$53~37~37 & 5.363 & 0.614 &          \\
120690 & 13~51~20.3 & $-$24~23~25 & 6.435 & 0.703 &          & 212330 & 22~24~56.4 & $-$57~47~51 & 5.310 & 0.665 &          \\
122742 & 14~03~32.4 & $+$10~47~12 & 6.288 & 0.733 &          & 214953 & 22~42~36.9 & $-$47~12~39 & 5.988 & 0.584 &          \\
126053 & 14~23~15.3 & $+$01~14~30 & 6.268 & 0.639 &          & 216803 & 22~56~24.1 & $-$31~33~56 & 6.482 & 1.094 & $-$4.272 \\
126525 & 14~27~33.0 & $-$51~55~59 & 7.829 & 0.682 &          & 217014 & 22~57~28.0 & $+$20~46~08 & 5.469 & 0.666 &          \\
\hline
\end{tabular}
\label{tab1}
\end{center}
\end{table*}

The CTIO 1.5m telescope is now operated by a consortium of about a dozen 
partners, known as SMARTS\footnote{http://www.astro.yale.edu/smarts/} 
(Small and Moderate Aperture Research Telescope System). This consortium 
runs the telescope in queue mode, with observations collected by a trained 
technician and made available for download by the principal investigator. 
Every 4-6 weeks the technician cycles between two available instruments 
({\it RC~Spec} and a broad-band infrared imager). In practical terms, this 
means that {\it RC~Spec} is available for an average of 26 weeks per year. 
It is important to note that SMARTS operates the {\it only} southern 
telescope run in queue mode with an aperture and instrument that are 
appropriate for this project. Aside from a dedicated survey telescope like 
the one at Mount Wilson, SMARTS is the only option that makes a 
time-domain survey like this one feasible.

The SMARTS consortium is currently seeking a new partner for 2007, who can 
contribute at the level of \$50,000 per year in return for the equivalent 
of 40 nights of observing time. Operating in queue mode, this allocation 
will allow us to obtain exposures for the entire available B+N sample 
(around 40 stars visible on any given date) three times each week whenever 
{\it RC~Spec} is on the telescope. Over time, this sampling rate has 
proven sufficient to determine the rotation periods for many of the stars 
in the Mount Wilson survey \cite{noy84a}, and in some cases allows a 
measurement or limit on the degree of latitudinal differential rotation.


\section{Dissemination} 

The results of our survey, including raw and calibrated time-series 
measurements of the Ca HK emission for the entire B+N sample, will be made 
available without restriction through the project web-site. Updates will 
be made approximately once every two months, after each epoch of 
observation with {\it RC~Spec} through the SMARTS queue scheduling system, 
which cycles between two available instruments at 4-6 week intervals. At 
the end of the first two years, we will publish the initial results of the 
survey in a refereed journal to advertise the availability of the data to 
the entire community. We will also promote the inclusion of the data in 
all appropriate future Virtual Observatory projects.


\section*{Acknowledgments}

The National Center for Atmospheric Research is a federally funded 
research and development center sponsored by the National Science 
Foundation.


\bibliographystyle{unsrt}


\end{document}